\documentclass[a4paper,10pt]{karticle}
\usepackage{graphicx}

\renewcommand{\thesection}{\S\arabic{section}}

\title
{
Dripping Faucet Dynamics Clarified by \\
an Improved Mass-Spring Model
}

\author
{ 
Ken {\sc Kiyono}\thanks{E-mail: k\_kiyono@buss96.phys.metro-u.ac.jp} and Nobuko {\sc Fuchikami}
}

\date
{(\today)
}

\begin{document}
\sloppy
\maketitle
\begin{center}
\textit{
Department of Physics, Tokyo Metropolitan University, Tokyo 192-0397
}
\end{center}

\begin{abstract}
An improved mass-spring model for a dripping faucet is presented.
The model is constructed based on the numerical results which we recently obtained from fluid dynamical calculations. 
Both the fluid dynamical calculations and the present mass-spring model exhibit a variety of complex behavior including transition to chaos in good agreement with experiments. Further, the mass-spring model reveals fundamental dynamics inherent in the dripping faucet system. 
\end{abstract}

{\footnotesize
\begin{center}
\end{center}
}

\newcommand{\lsim}{\begin{minipage}{12pt}
\vspace{0pt}$
\,\stackrel{\textstyle <}{\sim}$\end{minipage}}
\newcommand{\gsim}{\begin{minipage}{12pt}
\vspace{0pt}$
\,\stackrel{\textstyle >}{\sim}$\end{minipage}}
\baselineskip 4.5mm

\section{Introduction}
\hspace{.6em}
Motion of dripping water from a faucet is a well-known example of chaotic dynamical systems easily seen in daily life. Until now, many experiments have been performed to measure dripping time intervals between successive drops\cite{shaw84,Martien85,nunez89,wu89,dreyer91,austin91,sgp94,pinto95,penna95,rocha96,silva97,pinto98,kn98,shoji}. A variety of dynamical behavior has been reported, such as periodic and nonperiodic attractors, period doubling, intermittency, boundary and interior crises\cite{sgp94,pinto98}, Hopf bifurcation\cite{pinto95}, multiple stability\cite{wu89} and hysteresis\cite{sgp94}. 
Bifurcation is induced by a very small change of the flow rate, which is a main control parameter of the system. 
Return map of the dripping time intervals shows one-dimensional structure like tangled strings, in other words, map function is multi-valued for flow rates in some ranges\cite{shaw84,Martien85,kn98}.
In addition, ``unit structure''\cite{fuchi99} which repeatedly appears as the flow rate is varied has been observed in a bifurcation diagram in a wide range of the flow rate\cite{austin91,kn98,shoji}.

Although experimental results have been accumulated, no unified explanation exists for the complex phenomena observed in the dripping faucet system. So far, most of theoretical analyses have been made based on simple dynamical models that are composed of a variable mass and a spring (mass-spring model). The original mass-spring model which was proposed by Shaw\cite{shaw84} is described by
\begin{eqnarray}
 \frac{{\rm d}}{{\rm d}t} \left( m \frac{{\rm d}z}{{\rm d}t} \right) &=& - k z - \gamma \frac{{\rm d}z}{{\rm d}t} + m g,  \label{eq:shaw} \\
\frac{{\rm d}m}{{\rm d}t} &=& Q = {\rm const.}, \label{eq:massi}
\end{eqnarray}
where $z$ is the position of the forming drop, $m$ is its mass, and $g$,$k$ and $\gamma$ are constant parameters. The mass increases linearly with time by eq.~(\ref{eq:massi}). When the position $z$ reaches a critical point, a part $\Delta m$ of the total mass breaks away in proportion to the velocity at that moment. 
Although the model, being solved using an analog computer, simulated some experimental attractors, the parameter values ($g$, $k$, $\gamma$), the flow rate $Q$ and conditions at the breakup moment of a drop were not presented explicitly. 

S\'anches-Oritiz and Salas-Brito\cite{sanchez95a,sanchez95b} numerically investigated the mass-spring model in some regions of the parameter space ($Q$, $g$, $h$), where $h$ is another constant characterizing the falling mass as $\Delta m = h m \dot{z}$.

Without solving eqs.~(\ref{eq:shaw}) and (\ref{eq:massi}) numerically, Austin\cite{austin91} assumed an approximate equation of a damped-oscillation form: 
\begin{equation}
z(t) = A t^{1/3} + B \sin\left( \omega t^{2/3} + \Omega _i\right){\rm e}^{- \gamma t}.
\end{equation}
He explained repeats of unit structure composed of period-one and period-two motion in a bifurcation diagram in terms of a feed back loop in the phase angle $\Omega _i$. 

An approximate equation of another damped-oscillation form was used by d'Innocenzo and Renna\cite{Inno98}. They obtained bifurcation diagrams including the period-doubling cascade to chaos from both the approximate equation and the numerical integration of eqs.~(\ref{eq:shaw}) and (\ref{eq:massi}).

Although the mass-spring models exhibit complex behavior including a transition to chaos, several essential issues seem to be unclear: Are assumptions included in the models realistic? How to choose the parameter values? Can the models basically explain the dynamical features of real dripping faucet systems in a wide range of flow rates?

Recently, Fuchikami \textit{et al}. proposed an algorithm of fluid dynamical calculations for a dripping faucet. The algorithm is based on Lagrangian description instead of Eulerian one. A drop, which is assumed to be axisymmetric, is decomposed into many sliced disks and the dynamics of the fluid is described in terms of time evolution equations for disks under the influence of gravity, surface tension and viscosity.
The numerical simulation succeeded in reproducing well not only the shape of drops, but also the complex dynamical behavior observed in experiments, including the unit structure of the bifurcation diagram. Moreover, the fluid dynamical calculations yield a lot of information on, for example, the spring constant, the breakup conditions etc., that is essential if one aims to construct a simplified model of mass-spring type.

In this paper, we introduce another mass-spring model, an improved one reconstructed from the knowledge of the fluid dynamical calculations. Our mass-spring model can systematically explain various aspects of the complex dynamical behavior of the dripping faucet system. It will be clarified how the repeats of the unit structure appear in bifurcation diagrams. It will also be explained how a multi-valued map of the dripping time intervals is obtained. 

We first present results of our fluid dynamical calculations in \S 2, based on which the improved mass-spring model is constructed in \S 3.
The last section will be devoted to a summary. 
\section{Results of the Fluid Dynamical Calculation}
\hspace{.6em}
\setcounter{equation}{0}
We present here the results of our fluid dynamical calculations (FDC) of the dripping faucet system for a faucet of 5 mm in diameter. The flow rates are in the range $0.19 ~{\rm cm}^3 / {\rm s} \lsim Q$ $\lsim 0.35 ~ {\rm cm}^3 / {\rm s}$, which correspond to drop rates in the range from $2.2$ drops/s up to $6.0$ drops/s. 

We have chosen the length, time, mass, and pressure units as
\begin{equation}
l_0 \equiv \sqrt{\Gamma / \rho g}\,, \quad
t_0 \equiv (\Gamma/ \rho g^3)^{1/4} \,, \quad
m_0 \equiv \rho l_0^3\,, \quad
P_0 \equiv \sqrt{\rho g \Gamma}\,. 
\label{eq:unit} 
\end{equation}
For water at 20~$^{\circ}$C, $l_0 =0.27~$cm, $t_0=0.017~$s, $m_0 = 0.020~$g and $P_0 = $270~dyn/cm$^2$.  The faucet radius was fixed at $a = 0.916$ ($=2.5$ mm). 
The unit of viscosity was chosen as $\eta_0 \equiv 
(\rho \Gamma^3/g)^{1/4}$. Then viscosity 
is $\eta=0.002$ for water at 20 $^{\circ}$C. 

A control parameter is the flow velocity $v_0$ kept constant at the end of the faucet(=at the top of the fluid), which is related to the flow rate $Q$ as $Q=\pi a^2 v_0$. 
A breakup parameter $\sigma_{\mathrm{crit}}$ was introduced in the simulation. When the minimum cross-section of the necking region, $\sigma_{\mathrm{min}} \equiv \pi r_{\mathrm{min}}^2 $, reaches this value, we separated the drop from the remaining part and continued the simulation for the residue. (Another breakup parameter $\epsilon$ defined in the previous paper relates to $\sigma_{\mathrm{crit}}$ as $\epsilon = \sigma_{\mathrm{crit}}/\pi a^2$.)
For saving the computational time, a relatively large value was taken for the breakup parameter for long-term simulations\cite{fuchi99}, namely, $\sigma_{\mathrm{crit}}$ was set to be $10^{-2}$ (i.e., $\epsilon \approx 4 \times 10^{-3} $).  


\subsection{Fluid motion of a dripping faucet}   \label{ss21}
\hspace{.6em}
To construct a simplified model, it is important to observe the fluid motion obtained from FDC carefully.

\begin{figure}[htbp]
	\begin{center}
	\includegraphics[width=0.75\linewidth]{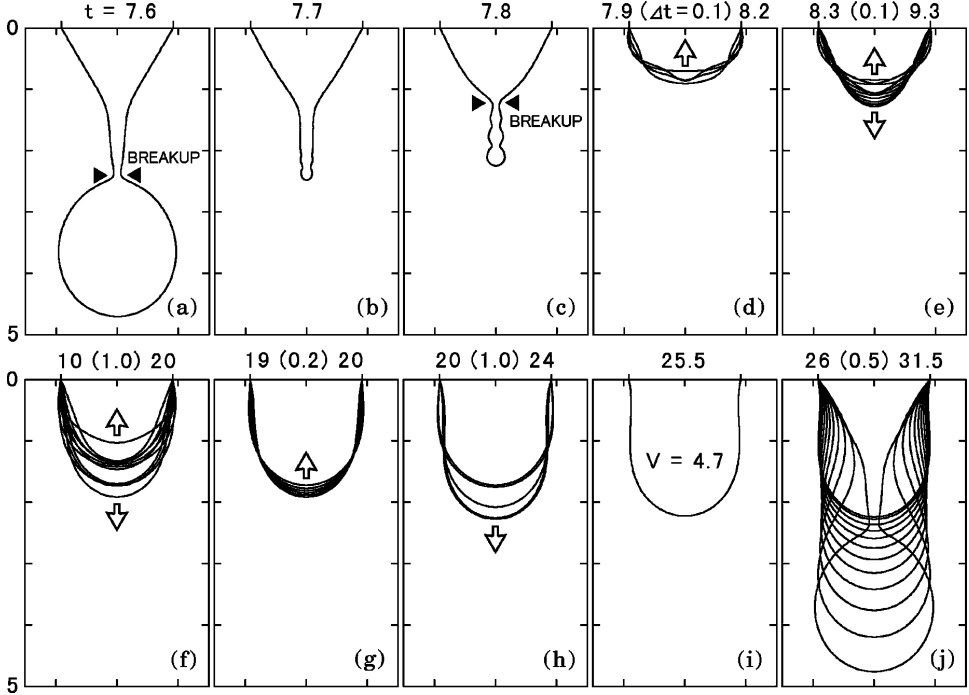}
	\end{center}
	\caption{The fluid dynamical calculations (FDC). Time evolution of water drops for $v_0 = 0.07$. }
    \label{fig:break}
\end{figure}

Deformation of water drops is presented in Fig.~\ref{fig:break} for the flow velocity $v_0 = 0.07$ and for the breakup parameter $\sigma_{\mathrm{crit}} = 10 ^{-2}$. 
Shape at a breakup moment is shown in Fig.~\ref{fig:break}(a). The breakup point of main drop appears at the lower root of a thin columnar bridge between the pendent mass (the conical part just below the faucet) and the nascent drop (the spherical part). Immediately after the separation of the main drop the thin columnar bridge recoils as the tip is formed like a round knob (Fig.~\ref{fig:break}(b)) and the secondary droplet, referred to as the satellite drop, is pinched off within a time $\Delta t \approx 0.2$ (Fig.~\ref{fig:break}(c)). The volume of a satellite drop is typically smaller than 1 \% of the largest drops. 
Just after the breakup moment, the fluid is stretched downward, so that the surface tension works as a restoring force. This causes oscillation of the fluid (Fig.~\ref{fig:break}(d)-(h)). 
When the increasing volume $V$ of the fluid amounts to the static critical volume $V_{\mathrm{crit}}$ ($\approx 4.7$ for the faucet radius $a = 0.916$), 
the next process, ``necking'', begins, because the fluid with the volume larger than $V_{\mathrm{crit}}$ cannot be suspended\cite{fuchi99}. 
In this process, a thin columnar bridge (between the pendant mass and the nascent drop) grows till the next breakup moment at which a round drop is separated from the remaining part.

In summery, a breakup of a drop occurs through two main processes, the oscillating process and the necking one. Breakup of a small satellite drop might be neglected in a simplified model, because that occurs in a relatively short time $\Delta t \approx 0.2$ after the breakup of a main drop. (As mentioned in the previous paper\cite{fuchi99}, the size ratio of a satellite drop to the main drop becomes smaller for smaller faucet radii.) Dripping of a drop repeats through these two processes appearing alternatively.
Figure \ref{fig:break} is a typical example for relatively slow flow velocities, i.e., for $0<v_0<0.16$ when $a = 0.916$. In all numerical results appearing in the present paper, the above basic mechanism of breakup commonly works.

\begin{figure}[hbtp]
	\begin{center}
	\includegraphics[width=0.52\linewidth]{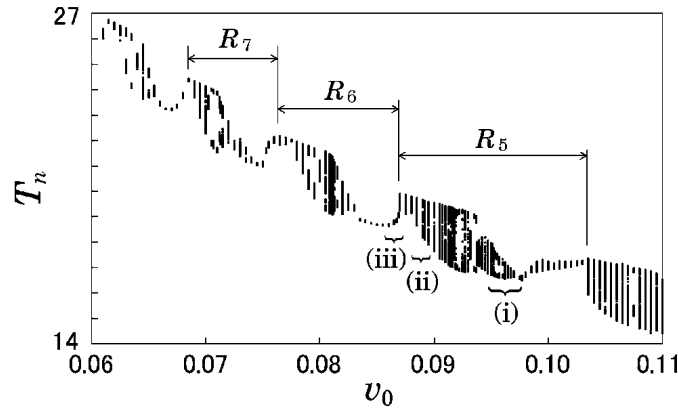}
	\end{center}
	\caption{Bifurcation diagram obtained from FDC. Plot of the dripping interval $T_n$ vs.~$v_0$．}
    \label{fig:bif}
\end{figure}

\subsection{Unit structure of bifurcation diagram} \label{ss22}
\hspace{.6em}
Figure \ref{fig:bif} is a bifurcation diagram obtained from the simulation data of the dripping time interval $T_n$, i.e., the time difference between the $(n+1)$-th and $n$-th breakup moments. We have ignored satellite drops that are smaller than 1 \% of the largest drops. 

In Fig.~\ref{fig:bif}, Period-one motion (P1), i.e., the motion in which the $T_n$ value is unique, is seen repeatedly, for example, at $v_0=0.0740$ and $v_0=0.0855$. In a region between these two $v_0$ values, for example, $0.75 < v_0 < 0.085$, the $T_n$ value is distributed over a finite range. This is seen in the form of a block, which repeats as $v_0$ increases. In the previous paper, we called this pattern unit structure. Looking at the bifurcation diagram with reference to $T_n$, one sees each unit structure occurs periodically, namely, at almost the same interval in $T_n$. A similar periodicity is observed experimentally\cite{kn98}. 

\begin{figure}[t]
	\begin{center}
	\includegraphics[width=.9\linewidth]{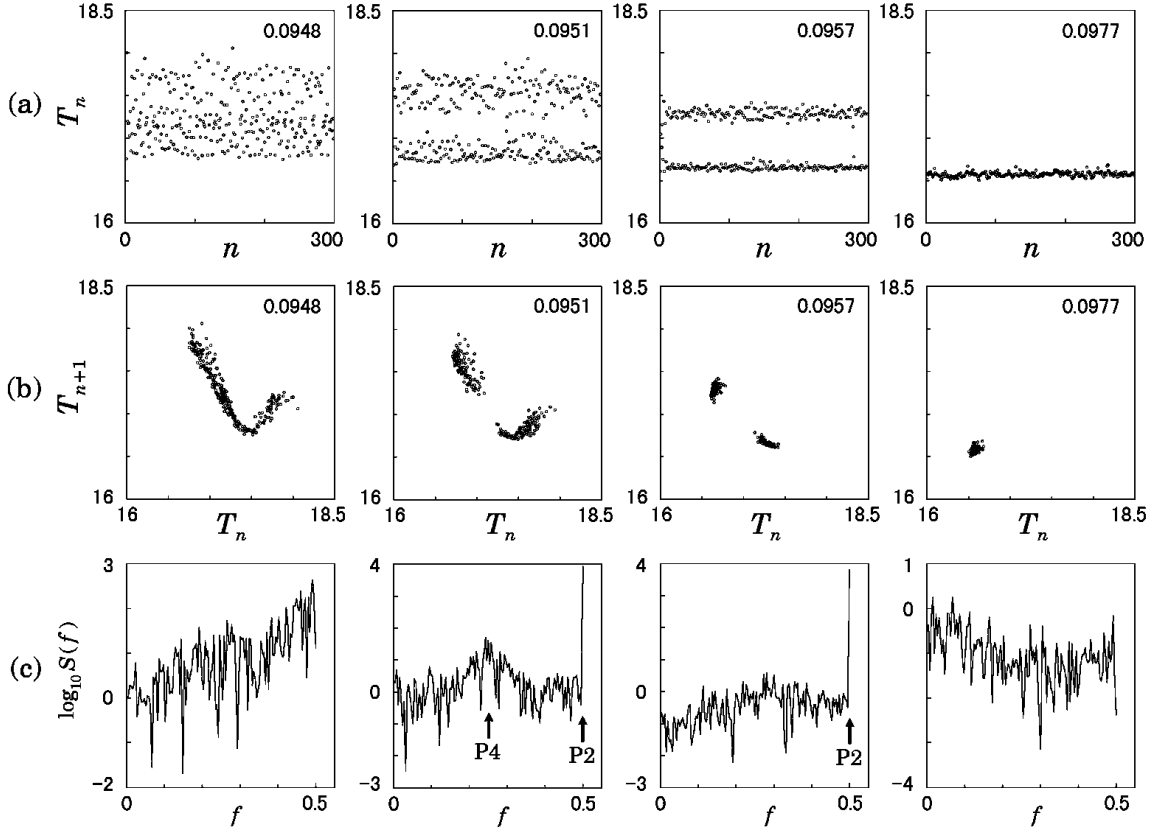}
	\end{center}
	\caption{FDC. (a) Plot of dripping time interval $T_n$ vs.~$n$ for various
values of $v_0$. Value of $v_0$ is written in each frame.
(b) Return map: plot of $T_{n+1}$ vs.~$T_n$.
(c) Semi-log plot of power spectrum of (a) obtained from
$2^{8}$ data points. }
    \label{fig:doubling}
\end{figure}

Plot of $T_n$ vs.~$n$, its return map (i.e., plot of $T_{n+1}$ vs.~$T_n$) and power spectrum are given in Fig.~\ref{fig:doubling} for several values of $v_0$ belonging to the same block, i.e., the same unit, whose parameter region is denoted as $R_5$. 

Period-one motion at $v_0=0.0997$ period-doubles backward to P2 motion at $v_0=0.0957$. The latter further period-doubles backward at $v_0=0.0951$, which can be seen as a peak at the frequency $f=1/4$ in the power spectrum of $T_n$. These results, together with the results for $v_0=0.0948$ in Fig.~\ref{fig:doubling}, suggest a backward period-doubling route to chaos. In each unit in Fig.~\ref{fig:bif}, backward period-doubling from P1 to P2 (as $v_0$ decreases) appears. In other words, the backward bifurcation from P1 to P2 occurs repeatedly with very small intervals of $v_0$, and in each interval a series of backward period-doubling cascade to chaos may occur. 
This suggests that experiments of measuring the dripping time intervals can be greatly influenced by a very small uncontrolled change of the flow rate. 
Also it might be possible that some experimental data of P1, P2 and P4 motion which arranged as if they were a part of a series of period-doubling cascade may belong to different series of cascade. 

\subsection{Relation between the unit structure and oscillation of fluid} \label{ss:osci}
\hspace{.6em}
We will show that the repeating unit structure in the bifurcation diagram is closely related to the oscillation of fluid. After a breakup, the position $z_{\mathrm{G}}$, the center of gravity of fluid, oscillates like a damped oscillator. 

\begin{figure}[hbtp]
	\begin{center}
	\includegraphics[width=0.53\linewidth]{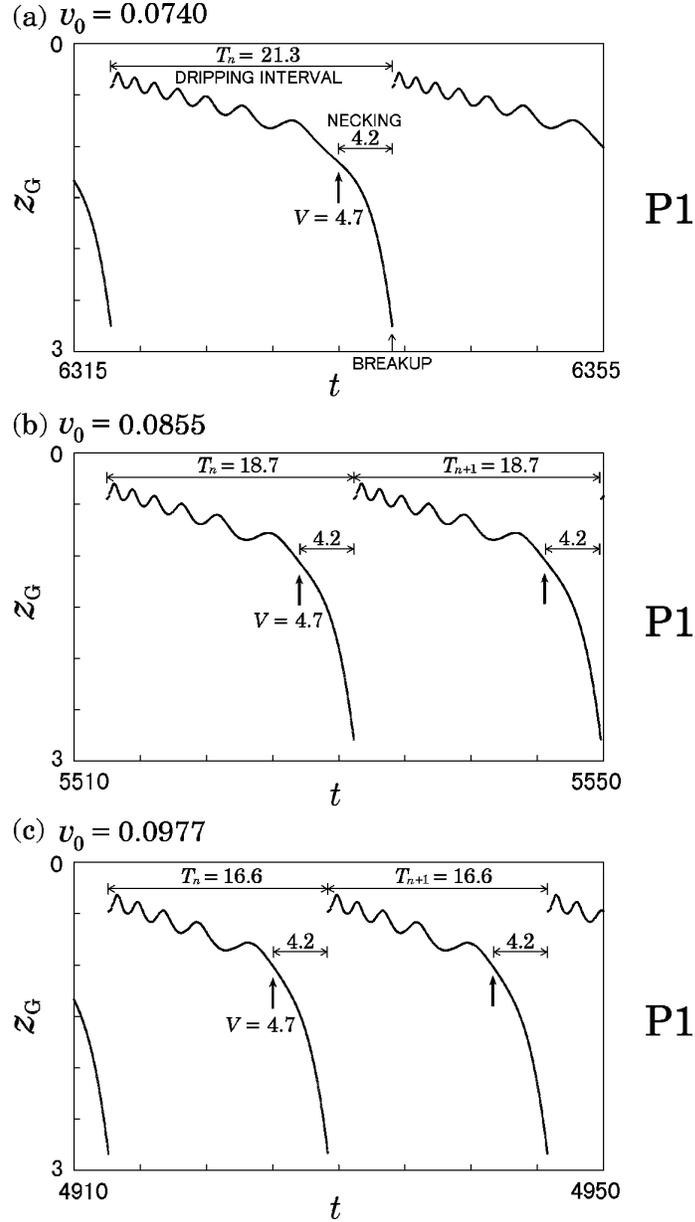}
	\end{center}
	\caption{FDC. Oscillation of the center of gravity of the fluid. Period-1 motion. Arrows indicate the moments at which $V=V_{\mathrm{crit}}=4.7$. }
    \label{fig:p1g}
\end{figure}

Let us observe first P1 motion in detail.
Figure \ref{fig:p1g} shows the oscillations of $z_{\mathrm{G}}$ for three kinds of P1 motion for $v_0 = 0.0740$, $0.0855$ and $0.0977$.
Corresponding to these $v_0$ values, the periods (= the dripping time intervals) are $T=21.3$, $18.7$ and $16.6$, which belong to three different units, whose regions in $v_0$ direction are $R_7$, $R_6$ and $R_5$, respectively. 
As mentioned in subsection \ref{ss21}, a breakup of a drop occurs through two main processes, oscillation and necking, which can well be observed in Fig.~\ref{fig:p1g}. Discontinuous changes in $z_{\mathrm{G}}$ seen in Fig.~\ref{fig:p1g} occur at breakup moments, at which a part of the fluid is lost. Arrows in Fig.~\ref{fig:p1g} indicate the points at which the total volume $V$ reaches $4.7=V_{\mathrm{crit}}$ that is the static critical value. Let $N$ be the oscillation frequency of $z_{\mathrm{G}}$ (= the number of peaks in the plot of $z_{\mathrm{G}}$ vs.~$t$) during a dripping time interval. In Fig.~\ref{fig:p1g}, one sees $N$ is 7, 6, and 5, corresponding to which the regions of units are $R_7$, $R_6$ and $R_5$, respectively. 
In this way, it was found that each unit is characterized by an integer $N$ such that the unit includes P1 motion whose oscillation frequency during a dripping time interval is $N$. 

After the volume exceeds the critical value, the fluid cannot be suspended and $z_{\mathrm{G}}$ drops rapidly because of necking. 
Time of the necking process, $\Delta t_{\mathrm{neck}}$, is not negligible in the sense that the total volume still increases for this time even after it reaches $V_{\mathrm{crit}}$. In Fig.~\ref{fig:p1g}, $\Delta t_{\mathrm{neck}}$, is almost equal to or more than 20 \% of the total dripping time interval. 
For $v_0 = 0.0977$, the total volume at the breakup moment amounts to $V_{\mathrm{total}} \approx 5.8$, about 27 \% of which is left after breakup. 
The necking time $\Delta t_{\mathrm{neck}}$ can change depending on details of the oscillation of the fluid, which causes variation of dripping time intervals. 
As an example, let us discuss the following P2 motion.

\begin{figure}[h]
	\begin{center}
	\includegraphics[width=0.68\linewidth]{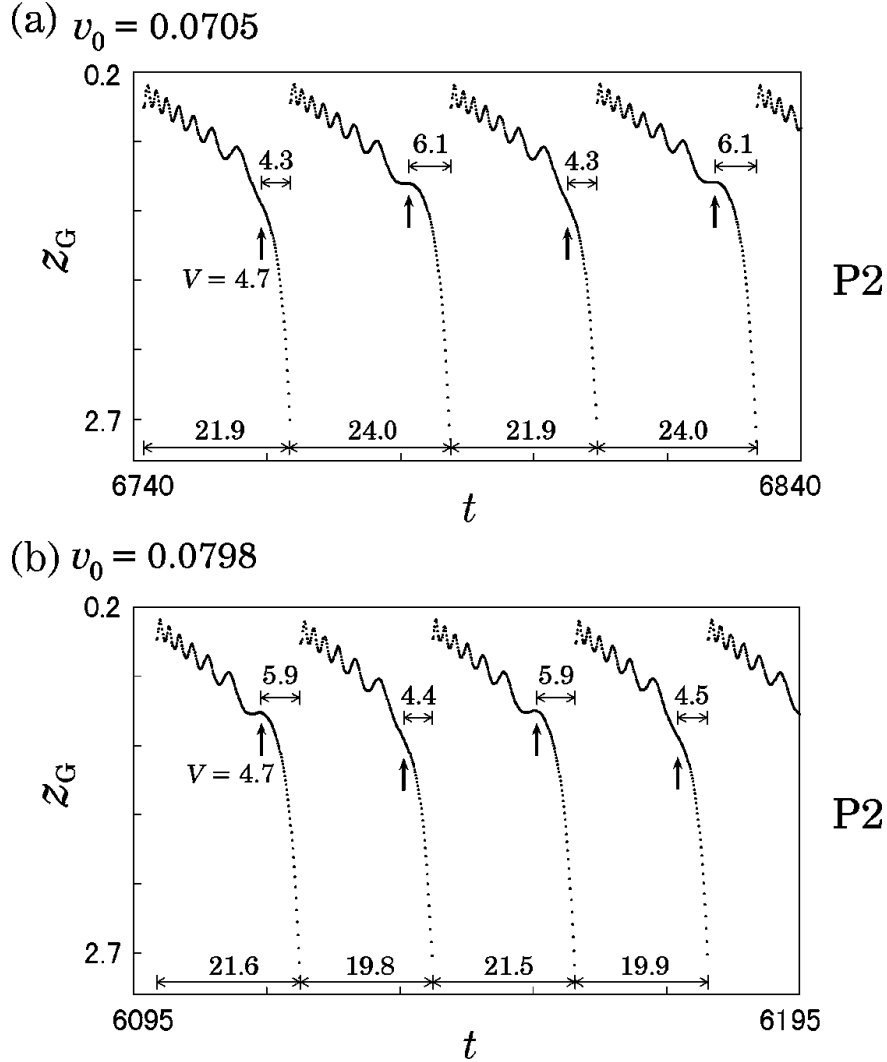}
	\end{center}
	\caption{FDC. Same as in Fig.~\ref{fig:p1g}. Period-2 motion. }
    \label{fig:p2g}
\end{figure}

In Fig.~\ref{fig:p2g}, P2 motion is plotted for $v_0=0.0705$ and $v_0=0.0798$, which belong to two different units in Fig.~\ref{fig:bif} specified by the intervals $R_7$ and $R_6$, respectively. For $v_0=0.0705$, oscillations with $N=7$ (the dripping time interval being $T_1 = 21.9$) and $N=8$ (the dripping interval being $T_2 = 24.0$) occur alternatively. The last (= the eighth) peak of $z_{\mathrm{G}}$ appears in every second time interval and is not so sharp as other peaks. 
Figure \ref{fig:p2g} shows that if the velocity $\dot{z}_{\mathrm{G}}$ at the moment of $V = 4.7 =V_{\mathrm{crit}}$ (indicated by arrows) is upward, an extra weak oscillation occurs before breakup. Observing the shapes of drops at various stages, we found that the waist of the fluid (at which $r=r_{\mathrm{min}}$) is more slender when $\dot{z}_{\mathrm{G}}$ at the moment of $V = V_{\mathrm{crit}}$ is downward than when it is upward, which causes a longer time interval $T_2$ than $T_1$.

In summary, each unit in the bifurcation diagram is specified by the symbol $R_i$ indicating a region of $v_0$. The suffix $i$ equals the minimum value of $N$, oscillation frequencies of the fluid during the time interval between two successive breakup moments.  

\begin{figure}[htp]
	\begin{center}
	\includegraphics[width=0.7\linewidth]{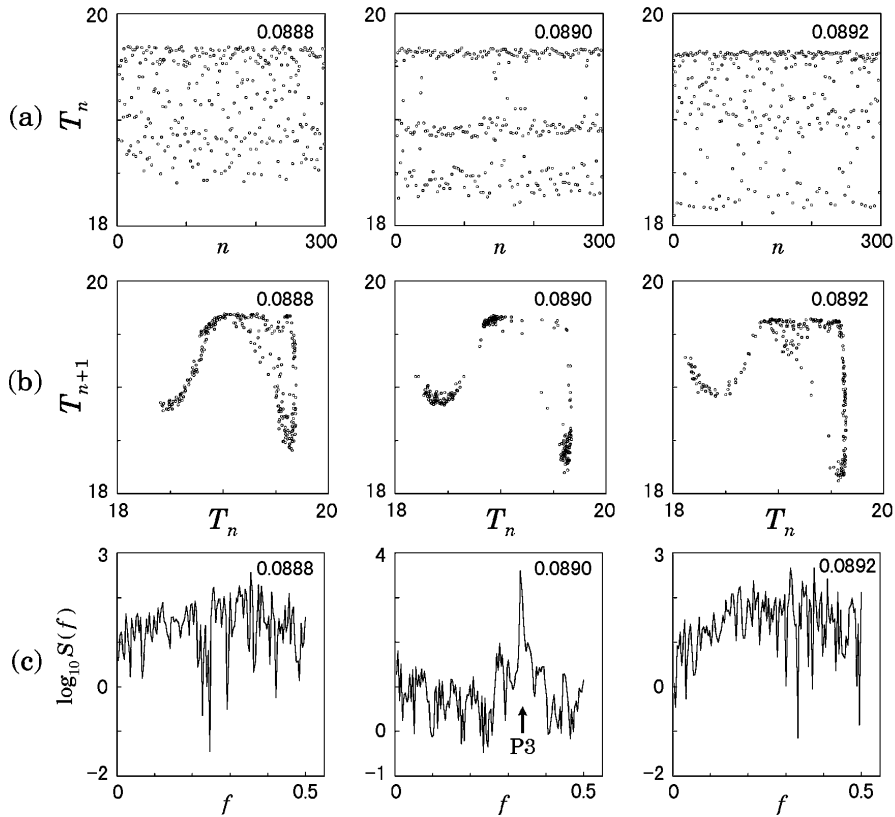}
	\end{center}
	\caption{FDC. Same as in Fig.~\ref{fig:doubling} but for different values of $v_0$. }
    \label{fig:p3}
\end{figure}

\subsection{Multi-valued return map}
\hspace{.6em}
Intermittent P3 motion is observed at $v_0=0.0890$, the spectrum of $T_n$ for this motion showing a peak at $f = 1/3$ (Fig.~\ref{fig:p3}). 
A small change of the $v_0$ value makes the spectrum broad, which indicates chaotic motion, for example, at $v_0 = 0.0888$ and $v_0 = 0.0892$. 
In Fig.~\ref{fig:p3}(b) we see that return maps for these $v_0$ values appear to be multi-valued, at least two-valued. 
This kind of multi-valued maps are often observed in experiments\cite{shaw84, kn98} as well as in simulations\cite{shaw84,sanchez95b}. 

\begin{figure}[th]
	\begin{center}
	\includegraphics[width=0.86\linewidth]{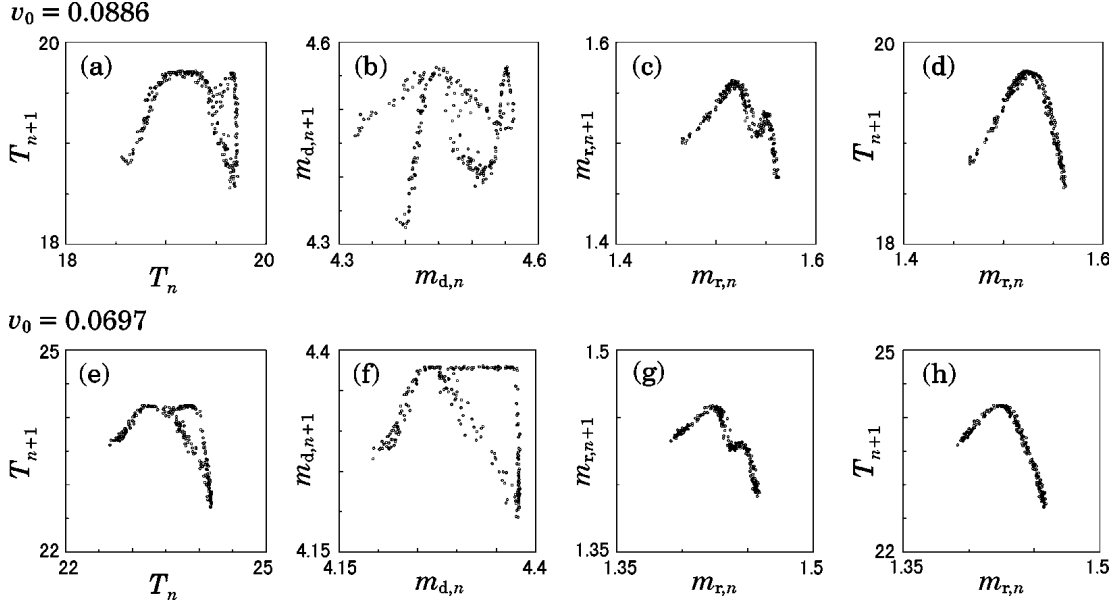}
	\end{center}
	\caption{Return maps obtained from FDC. The upper row for $v_0 = 0.0886$. The lower row for $v_0 = 0.0697$. (a),(e) Plot of $T_{n+1}$ vs.~$T_n$.
(b),(f) Plot of $m_{{\mathrm d},n+1}$ vs.~$m_{{\mathrm d},n}$ where $m_{{\mathrm d},n}$ represents the mass of the $n$-th falling drop. (c),(g) Plot of $m_{{\mathrm r},n+1}$ vs.~$m_{{\mathrm r},n}$ where $m_{{\mathrm r},n}$ is the $n$-th remnant mass. (d),(h) Plot of $T_{n+1}$ vs.~$m_{{\mathrm r},n}$. }
    \label{fig:maps}
\end{figure}

Figures \ref{fig:maps}(a) and \ref{fig:maps}(e) show other multi-valued return maps for $v_0=0.0886$ and $v_0 = 0.0697$. So far, we have plotted return map for dripping time intervals $T_n$, i.e., ($T_{n+1}$ vs.~$T_n$).
Note that experimental maps are usually plotted in the same way 
because measuring other quantities like mass is not easy in experiment. 
To investigate the mechanism of the multi-valued map, we plotted return maps using data of mass, where the mass is equal to the volume because $\rho = 1$. Figures \ref{fig:maps}(b) and \ref{fig:maps}(f) are plot for the dripping mass, i.e., ($m_{\mathrm{d},n+1}$ vs.~$m_{\mathrm{d},n}$).
They are again multi-valued maps. In contrast, the map function for the remnant mass, i.e., ($m_{\mathrm{r},n+1}$ vs.~$m_{\mathrm{r},n}$), is roughly single valued, although that is double humped (Figs.~\ref{fig:maps}(c) and \ref{fig:maps}(g)). 
Furthermore, if we plot the dripping time interval vs.~the previous remnant mass, i.e., ($T_{n+1}$ vs.~$m_{\mathrm{r},n}$), the return map becomes a simple unimodal function as seen in Figs.~\ref{fig:maps}(d) and \ref{fig:maps}(h). 
This implies that the time interval $T_{n+1}$ is directly determined only by the previous remnant mass $m_{\mathrm{r},n}$.

\begin{figure}[t]
	\begin{center}
	\includegraphics[width=0.60\linewidth]{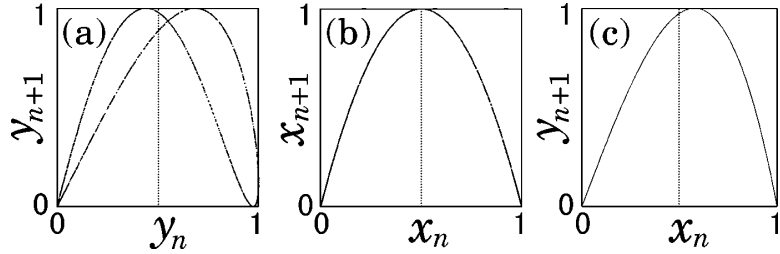}
	\end{center}
	\caption{A two-valued map (a) is obtained from a single-valued map (b) via a nonlinear asymmetric transformation (c) from $x_n$ to $y_{n+1}$. }
    \label{fig:logi}
\end{figure}

Relation between the single-valued map for $m_{\mathrm{r},n}$ and the two-valued map for $T_n$ is understood as follows. 
Let us consider two variables $x$ and $y$, and assume logistic map for $x$: 
\begin{equation}
 x_{n+1} = F (x_n) = 4 x_n (1-x_n).
\end{equation}
Further, assume that the next value of $y$ is uniquely determined by the present value of $x$ as
\begin{equation}
y_{n+1} = G(x_n) = \frac{3\sqrt{3}}{2} ( 1 - x_n ) x_n (1 + x_n )
\end{equation}
for  $0 \le x_n \le 1$ and $0 \le y_n \le 1$.
Then the map from $y_n$ to $y_{n+1}$ is given by
\begin{equation}
 y_{n+1} = G(x_n) = G(F(x_{n-1})) = G(F(G^{-1}(y_n))),
\end{equation}
which is two-valued as illustrated in Fig.~\ref{fig:logi}(a).
Note that the transformation $G(x)$ is asymmetric with respect to $x=1/2$.
This asymmetry causes the multi-valued map for $y_n$. 

The situation is roughly the same for the dripping faucet system. That is, $x_n$ and $y_n$ correspond to $m_{\mathrm{r},n}$ and $T_n$, respectively, which leads to a multi-valued return map for $T_n$. 

Some attractors of a dripping faucet look similar to H\'enon map as shown by Shaw\cite{shaw84}.
However, the above explanation suggests a difference in the fractal structure of attractor between the present system and H\'enon map. The difference arises because the present system is basically described in terms of a one-dimensional map, whereas H\'enon map is two-dimensional. We will discuss this point more detailedly in the next section. 

\begin{figure}[h]
	\begin{center}
	\includegraphics[width=.48\linewidth]{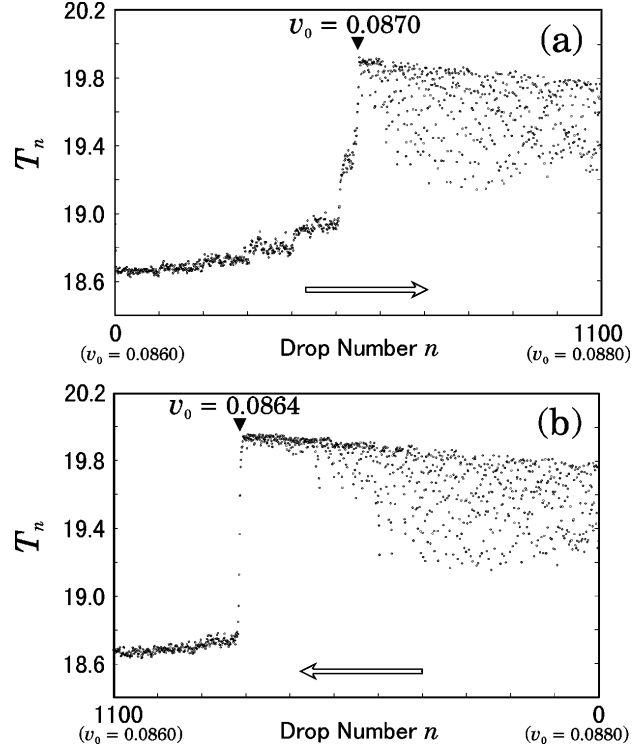}
	\end{center}
	\caption{Hysteresis observed in FDC. Plot of the dripping time interval $T_n$ for the $n$-th drop. (a) Increasing flow velocity. (b) Decreasing flow velocity. }
    \label{fig:hyst}
\end{figure}

\subsection{Hysteresis}
\hspace{.6em}
In Fig.~\ref{fig:hyst} is presented hysteresis: Different behavior is observed for some $v_0$ values depending on whether the $v_0$ value has been increasing (Fig.~\ref{fig:hyst}(a)) or decreasing (Fig.~\ref{fig:hyst}(b)). Hysteresis is observed also in an experiment by Sartorelli \textit{et al}.\cite{sgp94} In the next section, we will analyze our improved mass-spring model and show that the range of $v_0$ on which the hysteresis occurs is nearby the P1 motion belonging to the border of units.

\subsection{Mass dependence of the spring constant}
\hspace{.6em}
We have already seen that the fluid oscillates until the volume reaches the critical value $V_{\mathrm{crit}}$. This justifies the mass-spring model. However, for the model to be more realistic, one needs to know its spring constant. 

\begin{figure}[bh]
	\begin{center}
	\includegraphics[width=0.37\linewidth]{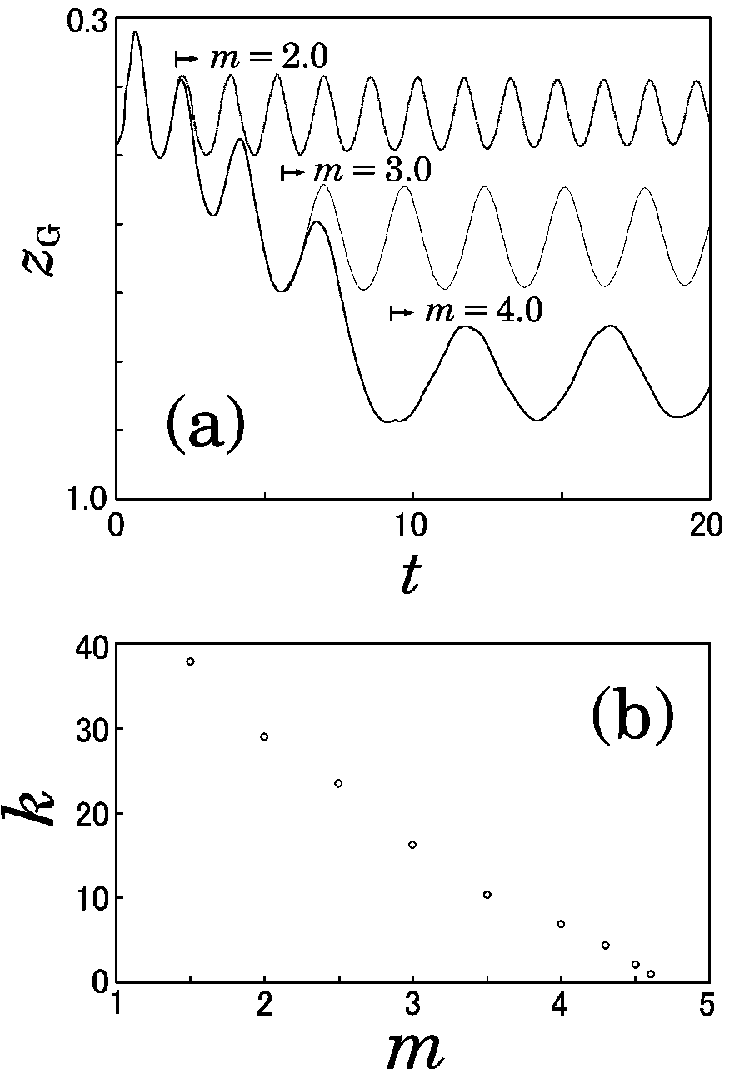}
	\end{center}
	\caption{FDC. (a) Oscillation of the center of gravity for several values of mass smaller than the critical value. (b) Mass dependence of spring constant. }
    \label{fig:spring}
\end{figure}

To estimate the spring constant $k$, we did a test simulation of FDC as follows: We started the simulation with some small initial volume ($V_0 = 1.4$) and some slow flow velocity ($v_0=0.07$). When the mass (= the volume) reaches a certain value $m$ for which we want to know the spring constant, we stopped the influx of water (i.e., $v_0$ was reset at zero). 
Then the fluid oscillates with very weak damping. Figure \ref{fig:spring}(a) shows the oscillation of the center of gravity of the fluid for several values of $m$. 
Measuring the period $T(m)$ of oscillation after the influx of water is stopped, the spring constant was obtained as $k(m) = m (2 \pi / T(m))^2$. The mass dependent spring constant thus obtained is plotted in Fig.~\ref{fig:spring}(b), which becomes zero near a critical value $m$ ($\approx V_{\mathrm{crit}}=4.7$). 
Based on these data, we will introduce a linear function $k(m)$ in our mass-spring model (eq.~(\ref{eq:kfunk}) in \ref{sub:mod}).

\subsection{Summary of the fluid dynamical calculations}
Several aspects observed in the dripping faucet system are summarized as follows:
\begin{enumerate}
	\item Breakup of a drop occurs through two processes, oscillation and necking. 
	\item Although the return map $T_{n+1}$ vs.~$T_n$ is a multi-valued function, the return maps $m_{{\mathrm r},n+1}$ vs.~$m_{{\mathrm r},n}$ and $T_{n+1}$ vs.~$m_{{\mathrm r},n}$ are single-valued.
	\item Repeating unit structure is observed in the bifurcation diagram over a wide range of the flow velocity. Each unit includes period-doubling and tangent bifurcation in a narrow range.
	\item Each unit is specified by the oscillation frequency of position $z_{\mathrm{G}}$ during the time interval $T_n$. 
	\item hysteresis.
\end{enumerate}

\section{Improved Mass-Spring Model} \label{sub:mod}
\hspace{.6em}
\setcounter{equation}{0}
We propose an improved mass-spring model based on the results of FDC for relatively low flow velocities $v_0<0.16$ (i.e., $Q < 0.35$~cm$^3$/s or equivalently, $6.0$~drops/s). 

We have observed $z_{\mathrm{G}}$, i.e., the movement of the center of gravity of the fluid in detail. Our mass-spring model describes only this degree of freedom explicitly like other mass-spring models.
It was found from the FDC that for low flow velocities breakup generally occurs through the process of ``regular'' necking. Here, ``regular'' means that the necking begins after but not before the total volume has reached the static critical value $V_{\mathrm{crit}}$. We take into account this issue in our mass-spring model. 


After breakup, the fluid starts to oscillate because of the surface tension which acts as a restoring force. The position $z_{\mathrm{G}}$ oscillates like a damped oscillator. Its mass-dependent spring constant has been obtained from the FDC. The dripping time interval strongly depends on the direction of the velocity $\dot{z}_{\mathrm{G}}$ at the moment at which $V=V_{\mathrm{crit}}$. If $\dot{z}_{\mathrm{G}}$ is upward, an extra oscillation of $z_{\mathrm{G}}$ occurs before the next breakup, which makes the dripping time interval longer. 

\subsection{Equation of motion}
\hspace{.6em}
Our mass-spring model obeys the following equations of motion: 
\begin{eqnarray}
 \frac{\mathrm{d}p}{\mathrm{d}t} &=& - k z - \gamma \frac{\mathrm{d}z}{\mathrm{d}t} + m g, \label{eq:damp} \\
 \frac{\mathrm{d}p}{\mathrm{d}t} &=& m \frac{\mathrm{d}^2 z}{\mathrm{d}t^2} + \left( \frac{\mathrm{d}z}{\mathrm{d}t} - v_0 \right) \frac{\mathrm{d}m}{\mathrm{d}t}, \label{eq:dp}
\end{eqnarray}
where $\gamma$ and $g$ are constant parameters. 
The units in eq.~(\ref{eq:unit}) have been used, so that $g = 1$. 
The damping parameter $\gamma$ was chosen as $\gamma = 0.05$. 
The second term in eq.~(\ref{eq:dp}) represents adhering of mass with a relative velocity $(\dot{z} - v_0)$. The mass increases with a constant flow rate $Q$:
\begin{equation}
\frac{\mathrm{d}m}{\mathrm{d}t} = Q = \mathrm{const.} \label{eq:mass}
\end{equation}
The term $v_0 \dot{m}$ ($=v_0 Q$) in eq.~(\ref{eq:dp}) was neglected because this is proportional to a small quantity $v_0^2$. (Note that $Q = \pi a^2 v_0$.)

Referring to the results of Fig.~\ref{fig:spring}(b), we assumed the mass-dependent spring constant $k$ takes the following linear form: 
\begin{equation}
k\left( m \right) =
\left\{ 
	\begin{array}{cl}
	 -11.4 m + 52.5 & ( m < 4.61 ) \\
	 0 & ( m \ge 4.61 ) 
	
	\end{array}
	\label{eq:kfunk}
\right. .
\end{equation}
When the mass $m$ amounts to $4.61$, the spring constant becomes zero, then the mass undergoes a free fall, which corresponds to the necking process of the fluid. An important difference of the present model from other mass-spring models is that the necking process is included explicitly.

We have made several assumptions for reset conditions at breakup moment. 
It is assumed that breakup occurs when the position $z$ reaches a critical point $z_{\mathrm{crit}}$. A part of the mass is then lost and the mass $m_{\mathrm{r}}$ is left. The remnant mass was renewed as
\begin{equation}
m_{\mathrm r} = 0.2m+0.3, \hspace{.5cm} {\rm when} \ z = z_{\mathrm{crit}},
\label{eq:reset1}
\end{equation}
where $m$ is the total mass at the breakup moment. 
The linear relation in eq.~(\ref{eq:reset1}) is oversimplified because this yields a single-valued map of $m_{\mathrm{d},n}$, which contradicts the FDC data (Figs.~\ref{fig:maps}(b) and \ref{fig:maps}(f)). If eq.~(\ref{eq:reset1}) is replaced by a nonlinear function of $m$, then the $m_{\mathrm{d},n}$ map also becomes multi-valued as well as $T_n$ map. We shall revisit this issue later on.

The position and velocity are renewed as
\begin{equation}
\left.
\begin{array}{lll}
z &=& z_0 = \mathrm{const}, \\
\dot{z} &=& 0, \\
\end{array}
\right\} \hspace{.5cm} {\rm when} \ z = z_{\mathrm{crit}}. 
\label{eq:reset2}
\end{equation}
The above reset conditions for $z$ and $\dot {z}$ are also simplified 
approximations. 
However, we found from the data of the FDC (see Figs. \ref{fig:p1g} 
and \ref{fig:p2g}) 
that these approximations are, at least qualitatively, acceptable.

\begin{figure}[h]
	\begin{center}
	\includegraphics[width=0.9\linewidth]{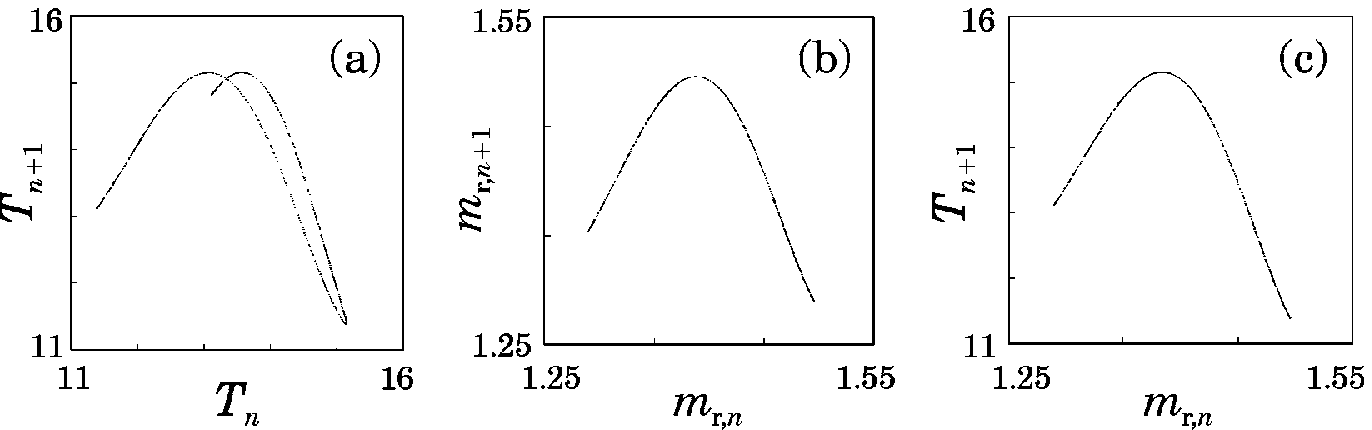}
	\end{center}
	\caption{MSM. $v_0 = 0.115$. (a) Map for the dripping time interval $T_n$, (b) for the remnant mass $m_{{\mathrm r},n}$. (c) Map from $m_{{\mathrm r},n}$ to $T_{n+1}$. }
    \label{fig:modmap1}
\end{figure}

Equations (\ref{eq:reset1}) and (\ref{eq:reset2}) imply that a set of the points on which the state is reset is located on a straight line in the phase space ($z$, $\dot{z}$, $m$), which will be shown in Fig.~\ref{fig:modphase} later. 
We chose $z_{\mathrm{crit}}=5.5$ and $z_0 = 2.0$. Note that the reset state depends only on the total mass just before the breakup. 
This implies the dripping time interval $T_{n+1}$ is decided uniquely, (by solving eqs.~(\ref{eq:damp}) and (\ref{eq:dp})) from the previous value $m_{\mathrm{r},n}$ of the remnant mass. 
Thus $m_{\mathrm{r},n+1}$ is uniquely decided only by $m_{\mathrm{r},n}$ from eq.~(\ref{eq:reset1}) because the total mass $m$ at the breakup moment is also decided uniquely by $m_{\mathrm{r},n}$ via $T_{n+1}$.
In this way, the present model is described in terms of a one-dimensional map from $m_{\mathrm{r},n}$ to $m_{\mathrm{r},n+1}$, which is consistent with data obtained from the FDC (Figs.~\ref{fig:maps}(c) and \ref{fig:maps}(g)).
Return maps of the present model will be presented later (Fig.~\ref{fig:modmap1}).

\begin{figure}[h]
	\begin{center}
	\includegraphics[width=.54\linewidth]{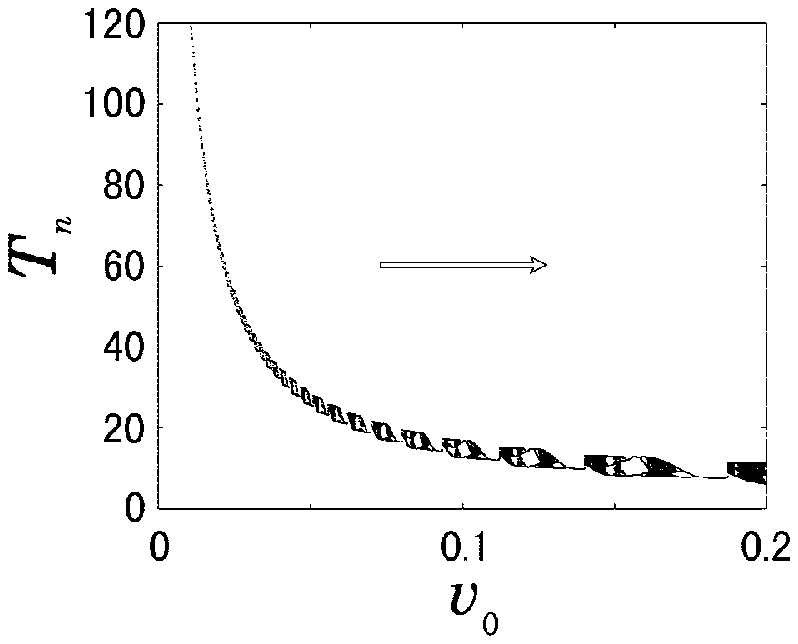}
	\end{center}
	\caption{Bifurcation diagram obtained from our mass-spring model (MSM) for a wide range of $v_0$. The diagram was made by raising $v_0$. The final state for the present $v_0$ value is chosen as the initial state for the next $v_0$ value. }
    \label{fig:widebif}
\end{figure}

\subsection{Bifurcation diagram}
\hspace{.6em}
We numerically solved eqs.~(\ref{eq:damp}) and (\ref{eq:dp}) together with eqs.~(\ref{eq:mass}) $\sim $ (\ref{eq:reset2}) and obtained a bifurcation diagram over a wide range of flow velocities $v_0$ (Fig.~\ref{fig:widebif}). We chose $a = 0.916$. The diagram was made by raising $v_0$. The final state at the present $v_0$ value is chosen as the initial state at the next $v_0$ value. 
The repeating unit structure in bifurcation diagrams which was observed in experiments\cite{austin91, kn98, shoji} and our FDC is reproduced. 

\begin{figure}[t]
	\begin{center}
	\includegraphics[width=.94\linewidth]{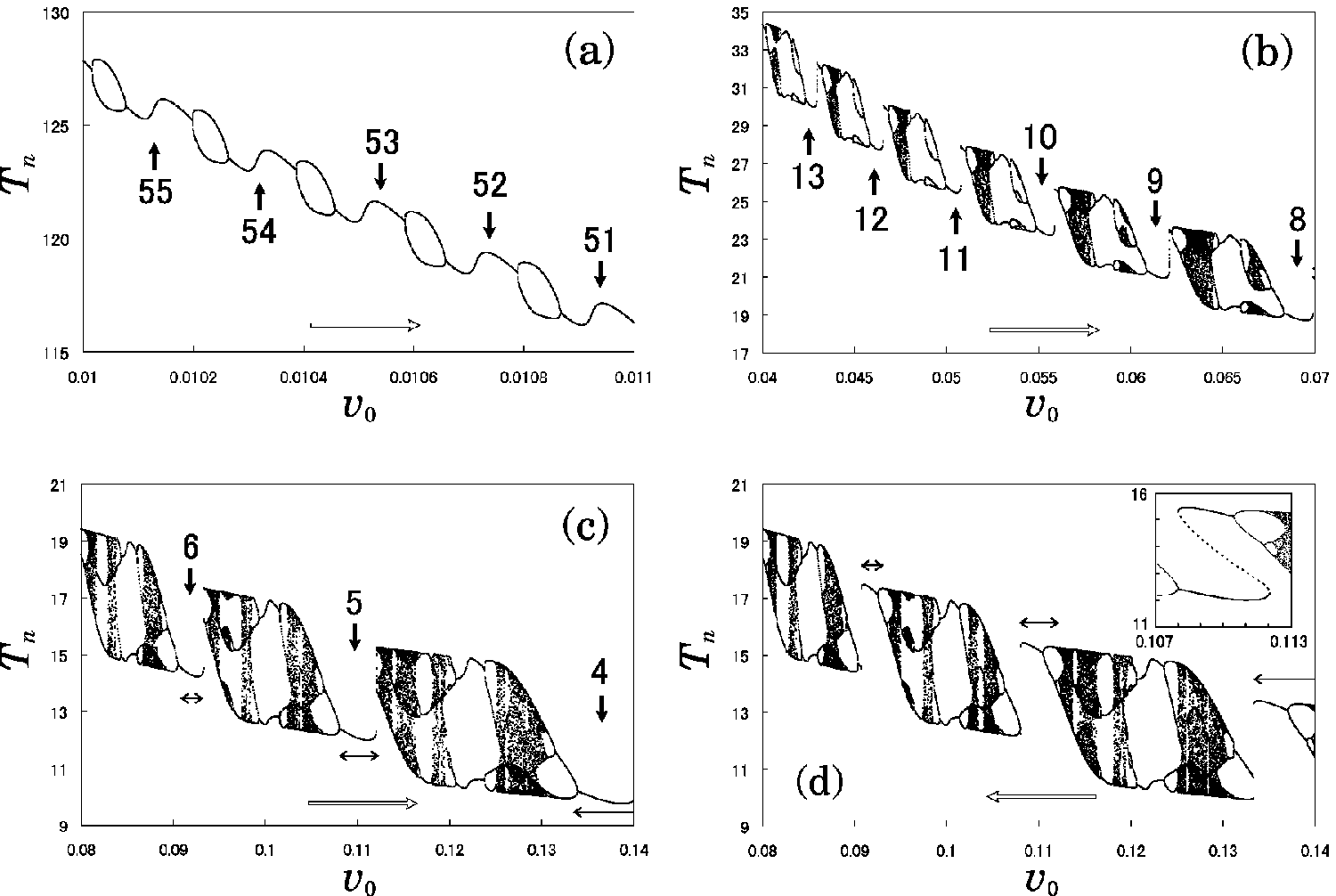}
	\end{center}
	\caption{MSM. Bifurcation diagrams. Enlargement of several parts of Fig.~\ref{fig:widebif}. The number in each frame represents the frequency $N$ of oscillations between two successive breakup moments. In (a), (b) and (c), the control parameter $v_0$ was raised. (d) is the same as (c) but $v_0$ was decreased. For (c) and (d), the bifurcation diagrams are different in some ranges of $v_0$ (indicated with $\leftrightarrow$), which shows hysteresis. The inset shows a hysteresis curve, in which a dotted line represents an unstable orbit. }
    \label{fig:3stage}
\end{figure}
\begin{figure}[ht]
	\begin{center}
	\includegraphics[width=.95\linewidth]{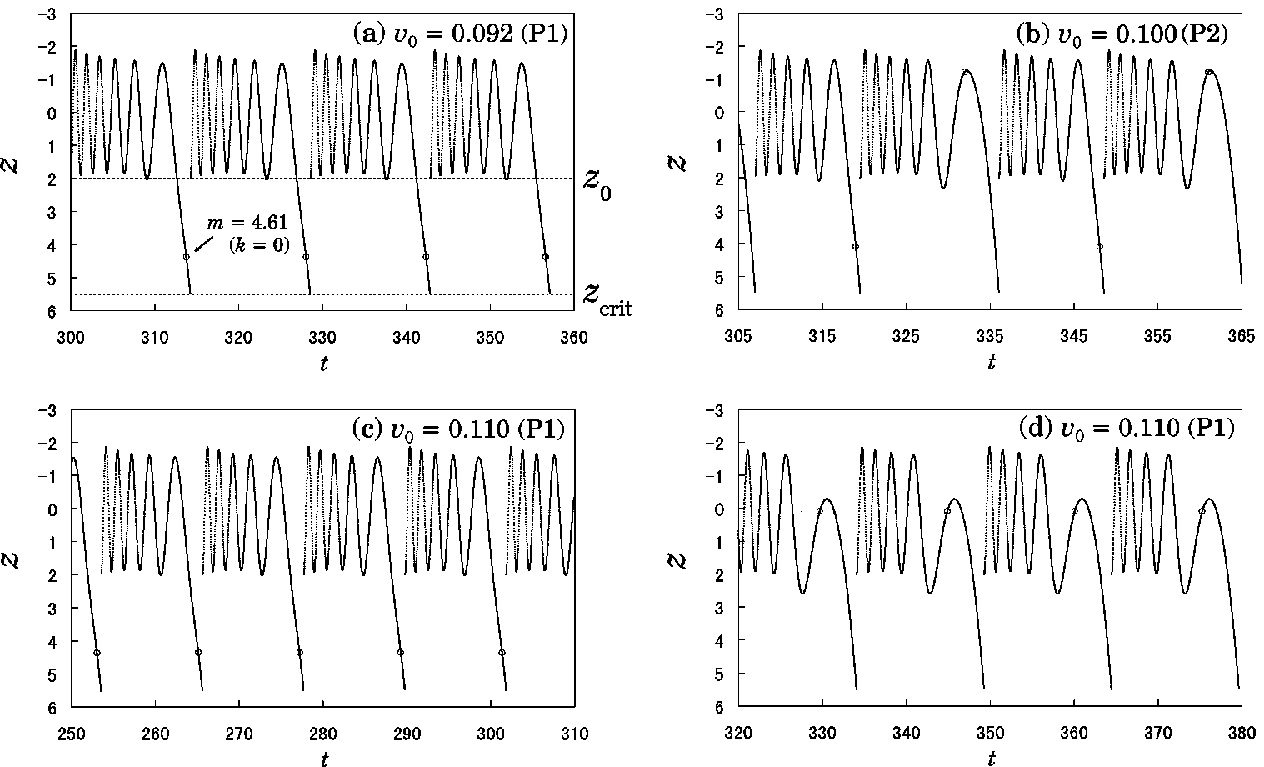}
	\end{center}
	\caption{Oscillation of the position $z$ in MSM. Circles indicate the moments at which the spring constant becomes zero. (a) P1 motion. (b) P2 motion. (c),(d) Different types of P1 motion at the same $v_0$ value: $v_0 = 0.110$.}
    \label{fig:modzg}
\end{figure}

In Fig.~\ref{fig:3stage}, bifurcation diagrams are shown for three different ranges of $v_0$ included in Fig.~\ref{fig:widebif}. 
In a range of very small $v_0$ (Fig.~\ref{fig:3stage}(a)), each unit is composed of only P1 and P2 motion. 
A similar type of unit structure is observed experimentally\cite{austin91, kn98, shoji}. We have not obtained this type of unit structure from FDC because long-term simulations for such small flow velocities take too long computational time to obtain enough data at many $v_0$ values. 

For higher flow velocities (Fig.~\ref{fig:3stage}(b)), each unit includes forward and backward period-doubling cascades that start from P2 motion, in addition to the normal (i.e., starting from P1 motion) period-doubling cascade to chaos. Similar data are also obtained in experiments\cite{austin91, wu89, kn98}. 
The numbers in Fig.~\ref{fig:3stage} denote the oscillation frequency $N$ during the dripping time interval $T_n$ for the P1 motion at the points indicated with allows. 
The frequency $N$ increases by one by moving to the neighboring unit on the left, i.e., the side of lower flow velocity. 

The oscillation of the position $z$ is represented in Fig.~\ref{fig:modzg}. Figures \ref{fig:modzg}(a) and \ref{fig:modzg}(c) show P1 motion at two $v_0$ values, $v_0 = 0.092$ and $0.110$, having $N=5$ and $6$ respectively. They belong to two neighboring units in Fig.~\ref{fig:3stage}(c). 
In P2 motion at $v_0 = 0.100$ (see Figs.~\ref{fig:3stage}(c) and \ref{fig:modzg}(b)), two different frequency $N=5$ and $6$ appear alternatively. 
These features have been observed in the FDC (Figs.~\ref{fig:p1g} and \ref{fig:p2g}). 

Further, the present model exhibits hysteresis as seen in Figs.~\ref{fig:3stage}(c) and \ref{fig:3stage}(d). Figures \ref{fig:modzg}(c) and \ref{fig:modzg}(d) show different types of P1 motion observed at the same $v_0$ value: $v_0 = 0.110$. 
They belong to the lower and upper branch of hysteresis curve and obtained by raising and reducing $v_0$, respectively. 
In Fig.~\ref{fig:modzg}, circles indicate the moment at which the mass reaches the critical value $m=4.61$ and the spring constant $k$ becomes zero. 
Comparing Figs.~\ref{fig:modzg}(c) and \ref{fig:modzg}(d), we see that the dripping time interval $T_n$ is longer in (d) mainly because the velocity $\dot{z}$ is upward at the moment of $k=0$. 
The hysteresis will be discussed in more detail later on (in the next subsection) together with map functions. 

The repeating unit structure of the bifurcation diagram shows a scaling law, which has been found experimentally\cite{austin91, kn98, shoji}. As was mentioned in subsection \ref{ss22}, each unit occurs periodically, i.e., at almost the same interval in $T_n$. 
Let us specify bifurcation points of forward period-doubling from P1 to P2 as ($v_{0,N}$, $T_{n,N}$), where $N$ is the oscillation frequency. Then these points depend on $N$ as
\begin{eqnarray}
T_{n,N} & \approx & 2.2 N + 3.9, \label{eq:scale1}\\
v_{0,N} & \approx & 0.58 N^{-1.0}. \label{eq:scale2}
\end{eqnarray}
Equation (\ref{eq:scale1}) corresponds to the periodicity of the repeating unit structure in $T_n$ direction. The scaling law, eq.~(\ref{eq:scale2}), agrees well with the experimental scaling law $Q \propto N^{-1.1}$ obtained by Katsuyama and Nagata\cite{knpr}. 
An empirical scaling law which agrees with the above for sufficiently large $N$ was also obtained by Shoji\cite{shoji}. 

\subsection{Return map} \label{sub:rmap}
\hspace{.6em}
The present mass spring model is basically described by a one-dimensional map. Once $m_{\mathrm{r},n}$ (the $n$-th remnant mass just after breakup) is given, $T_{n+1}$ (the next dripping time interval) is decided uniquely by eqs.~(\ref{eq:damp}) and (\ref{eq:dp}). 
The map from $m_{\mathrm{r},n}$ to $T_{n+1}$, $T_{n+1} = G(m_{\mathrm{r},n})$, is shown in Fig.~\ref{fig:modmap1}(c). Then the next value of the remnant mass, $m_{\mathrm{r},n+1}$, is uniquely determined as
\begin{equation}
m_{\mathrm{r},n+1} = 0.2 \left(m_{\mathrm{r},n} + Q G (m_{\mathrm{r},n}) \right) + 0.3, 
\end{equation}
because of eqs.~(\ref{eq:mass}) and (\ref{eq:reset1}). (Fig.~\ref{fig:modmap1}(b).) 
Therefore, the map from $T_n$ to $T_{n+1}$ is also one-dimensional even if it can be multi-valued. 
(See Fig.~\ref{fig:modmap1}(a).) 
We have illustrated how the map can be multi-valued in Fig.~\ref{fig:logi}. 
A similar return map of $T_{n+1}$ vs.~$T_n$ which looks like tangled strings was also observed in the FDC (Figs.~\ref{fig:maps}(a) and (e)). 
This suggests the possibility that experimentally observed return maps of $T_n$ can also be one-dimensional, even when the maps look like complicatedly tangled strings. 

\begin{figure}[htp]
	\begin{center}
	\includegraphics[width=.6\linewidth]{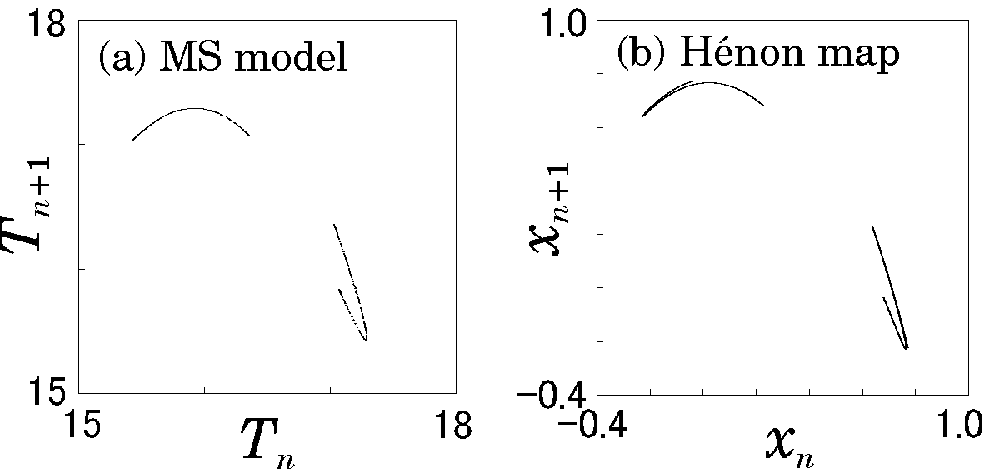}
	\end{center}
	\caption{(a) MSM. Return map of $T_{n}$ for $v_0=0.094$. (b) H\'enon map: $x_{n+1}= -0.33 y_n + 1 - 2.07 x_n^2$, $y_{n+1} = x_n$. }
    \label{fig:henon}
\end{figure}

The present mass-spring model is thus basically different with H\'enon map in the structure of chaotic attractor. This is because H\'enon map is two-dimensional, i.e., the next value $x_{n+1}$ is decided not only by the present value $x_n$ but also by the previous value $x_{n-1}$:
\begin{equation}
x_{n+1} = 1 - a x_n^2 + b x_{n-1}. 
\end{equation}
In Fig.~\ref{fig:henon} \, some attractors are represented for the MSM and H\'enon map. At first sight, they look similar. However, H\'enon attractor includes selfsimilar structure on a two-dimensional plane, whereas the attractor of MSM is restricted on a one-dimensional manifold. 
Note that the situation does not change even if the linear reset condition (eq.~(\ref{eq:reset1})) is replaced by a nonlinear relation. The map function $m_{\mathrm{d},n+1}$ vs.~$m_{\mathrm{d},n}$ then becomes multi-valued as seen in Figs.~\ref{fig:maps}(b) and \ref{fig:maps}(f) of FDC but the dynamics is still described by a one-dimensional map, namely the next value $x_{n+1}$ is determined only by the present value $x_n$. 

We can also improve the reset condition $\dot{z}=0$ such that the renewed $\dot{z}$ value depends on the $\dot{z}$ value just before the breakup. 
Then the dimensionality of the system increases and the chaotic attractor may exhibit selfsimilar structure as H\'enon map. 
Such a refinement of the model is, however, not so important in the sense that the essential aspect, i.e., the return map which looks like tangled strings can be obtained  by a one-dimensional map. 

\begin{figure}[htpb]
	\begin{center}
	\includegraphics[width=.8\linewidth]{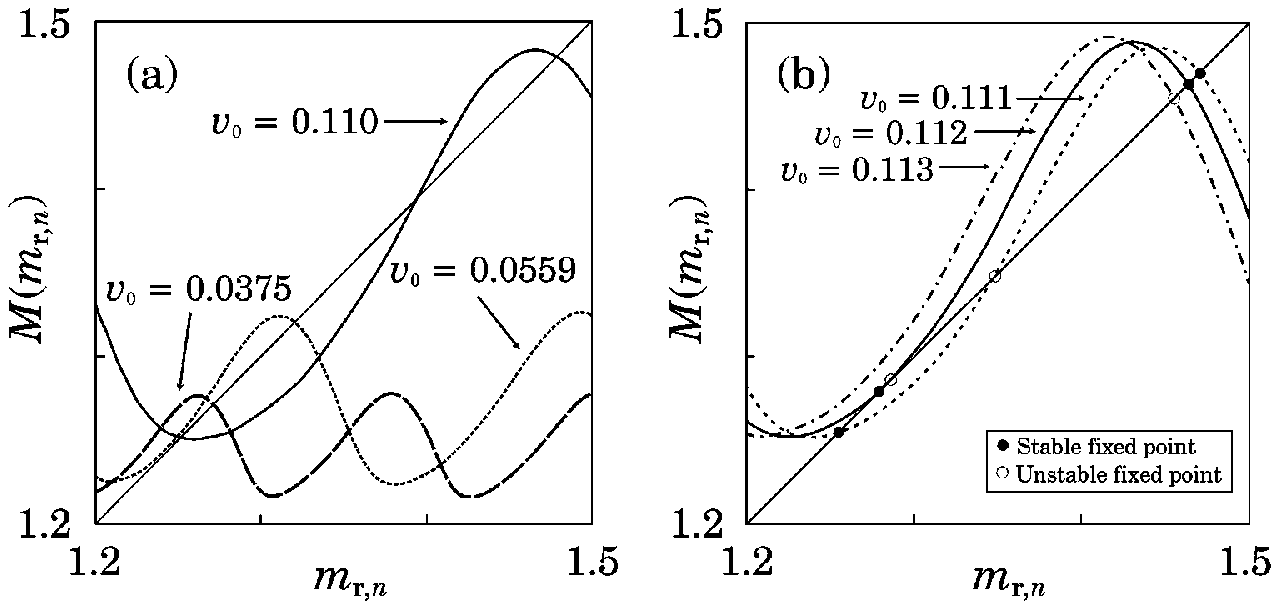}
	\end{center}
	\caption{MSM. Map function $m_{{\mathrm r},n+1} = M (m_{{\mathrm r},n}, v_0)$ for various values of $v_0$. }
    \label{fig:modmap2}
\end{figure}

Figure \ref{fig:modmap2} shows map function $m_{\mathrm{r},n+1} = M (m_{\mathrm{r},n}, v_0)$ for several values of the flow velocity $v_0$. 
The map $M$ is an oscillating function of $m_{\mathrm{r},n}$. 
The oscillation causes the repeats of unit structure.
The amplitude and the maximum slope of the map function becomes large with $v_0$. 
This is the reason why the unit structure is simple for small $v_0$ values and becomes more complicated as $v_0$ increases. 
In a range of small enough $v_0$ values, 
each unit includes P1 and P2 motion only. 
In a range of larger $v_0$ values, period-doubling bifurcation can occur more than once in the same unit. For large enough $v_0$ values, each unit includes forward and backward period-doubling cascade to chaos and tangent bifurcation. 

Hysteresis can be understood from Fig.~\ref{fig:modmap2}(b). 
There are two stable fixed points at, for instance, $v_0 = 0.111$. 
If this $v_0$ value has been reached by an increasing process of $v_0$, then the fixed point on the left (with a smaller $m_{\mathrm{r},n}$ value) is realized. As $v_0$ increases further, the fixed point on the left shifts to the right (see for $v_0 = 0.112$ in Fig.~\ref{fig:modmap2}(b)) until this fixed point disappears by forward tangent bifurcation. 
At this moment, the other fixed point (with a larger $m_{\mathrm{r},n}$ value) is still stable (at a $v_0$ value slightly larger than $0.112$) and this fixed point is realized, which causes discontinuous increase in the $m_{\mathrm{r},n}$ value. 
In an decreasing process of $v_0$, the fixed point on the right (with a larger $m_{\mathrm{r},n}$ value) is realized first, and when this fixed point disappears by backward tangent bifurcation, the other stable fixed point on the left is realized. The hysteresis occurs in this way. 

Some experimental data have been explained in terms of boundary 
crisis.\cite{sgp94}
However, as suggested from the bifurcation diagram, our mass-spring
model yields no boundary crisis.
Boundary crisis occurs for example in a quadratic map when an unstable
fixed point that is a basin boundary of a chaotic attractor collides with
an edge of this chaotic attractor.
In contrast, the unstable fixed point expressed by the dotted line in 
the inset of Fig. \ref{fig:3stage} (d) is a basin boundary of two fixed 
point attractors
and it disappears together with one of these fixed point attractors by
tangent bifurcation.

\begin{figure}[htpb]
	\begin{center}
	\includegraphics[width=.7\linewidth]{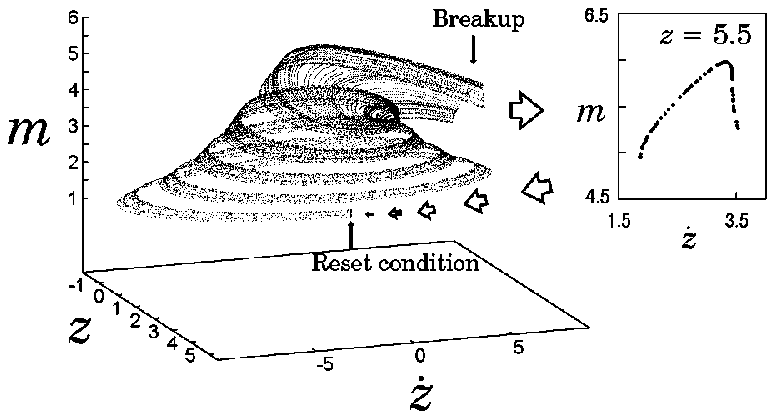}
	\end{center}
	\caption{Phase space portrait of MSM. $v_0 = 0.115$. The inset is a Poicar\'e surface of section at the critical position $z=z_{\mathrm{crit}} = 5.5$. }
    \label{fig:modphase}
\end{figure}

Finally we present chaotic trajectories in the phase space ($z$, $\dot{z}$, m) in Fig.~\ref{fig:modphase}. The state point is on the plane $z = z_{\mathrm{crit}}=5.5$ at each breakup moment.
A set of state points just after the breakup moments is on a line segment satisfying $z=2$ and $\dot{z}=0$ (reset conditions). This line segment is stretched and folded on the plane of $z = z_{\mathrm{crit}}$. Then the image again mapped onto the initial line segment by the reset conditions expressed as eqs.~(\ref{eq:reset1}) and (\ref{eq:reset2}). 

\section{Summary}
\hspace{.6em}
We introduced an improved mass-spring model based on the results of our fluid dynamical calculations. 
The FDC data reproduced the repeating unit structure in the bifurcation diagram, stable P2 motion included in the middle of each unit, forward and backward period-doubling starting from this P2 motion, hysteresis etc. which were observed experimentally. 
The FDC showed that the repeating unit structure is caused by the oscillation of fluid between two successive breakup moments. 
The FDC also showed that the necking process plays an important role in changing the dripping time interval. 

Our mass-spring model corresponds to a one-dimensional map system, i.e., the simplest system that can yield chaos. 
In spite of that, the model succeeded in reproducing systematically a variety of dynamical behavior observed in many experiments. 
Especially, the model supplemented a part of the bifurcation diagram in a range of very small flow velocities $v_0$ that was not obtained by FDC because of long computational time. 
For these small $v_0$ values, the model showed that each unit structure in the bifurcation diagram is composed only of P1 and P2 motion. 
For relatively large $v_0$ values, each unit includes period-doubling cascade to chaos. The period-doubling cascade starts not only from P1 motion but also from P2 in the middle of the unit. For the latter, the cascade occurs both forward and backward. 
These kinds of structure agree well with experimental observations. 
The two-valued map for the dripping time interval was explained from a single-valued map of the remnant mass. 
Mechanism of the hysteresis was understood based on the return map. 

The simple dynamics of the present mass-spring model clarified fundamental mechanisms inherent in the complex behavior of the dripping faucet system. 

\section*{Acknowledgement}
We are grateful to Dr.~T.~Katsuyama for valuable discussions. 
We thank Professor M.~Shoji for showing us his data prior to publication.

\vspace{1cm}

\end{document}